\documentclass[12pt,fleqn]{letter}
\usepackage[dvipdfm]{graphicx,color}
\begin{document}
\small
\newcommand{\beeq}{\begin{equation}}   \newcommand{\eneq}{\end{equation}}
\newcommand{\besubeq}{\begin{subequations}}  
\newcommand{\ensubeq}{\end{subequations}}
\newcommand{\befg}{\begin{figure}} \newcommand{\enfg}{\end{figure}}
\newcommand{\lbf}[1]{\label{ff:#1}} \newcommand{\rff}[1]{\ref{ff:#1}}

\newcommand{\lb}[1]{\label{eq:#1}} \newcommand{\rf}[1]{\ref{eq:#1}}
\newcommand{\vp}{\varpi} 
\newcommand{\dr}[2]{\frac{d#1}{d#2}}
\newcommand{\pl}{\partial}
\newcommand{\pldr}[2]{\frac{\partial#1}{\partial#2}}

\newcommand{\Om}{\Omega} \newcommand{\om}{\omega}
\newcommand{\al}{\alpha}
\newcommand{\OmF}{\Omega_{\rm F}}  \newcommand{\OmH}{\Omega_{\rm H}}

\newcommand{\vcjp}{\mbox{\boldmath $j$}_{\rm p}} 
\newcommand{\vcS}{\mbox{\boldmath $S$}}
\newcommand{\vcv}{\mbox{\boldmath $v$}}   \newcommand{\vcj}{\mbox{\boldmath $j$}}   
\newcommand{\vcvF}{\mbox{\boldmath $v$}_{\rm F}} 
\newcommand{\vcBp}{\mbox{\boldmath $B$}_{\rm p}} 
\newcommand{\Bp}{B_{\rm p}} 
\newcommand{\vcEp}{\mbox{\boldmath $E$}_{\rm p}}
\newcommand{\uvt}{\mbox{\boldmath $t$}}
\newcommand{\vcnb}{\mbox{\boldmath $\nabla$}} 
\newcommand{\jvl}{j_{\perp}} \newcommand{\jpl}{j_{\parallel}} 
 \newcommand{\rH}{r_{\rm H}}  \newcommand{\SB}{S$_{\rm B}$}
\newcommand{\SA}{S$_{\rm A}$} \newcommand{\SF}{S$_{\rm F}$} 
 \newcommand{\Sinf}{S$_{\infty}$}  
\newcommand{\Sffinf}{S$_{{\rm ff}\infty}$} 
\newcommand{\Spffinf}{S$_{{\rm pff}\infty}$}
\newcommand{\SpffH}{S$_{{\rm pffH}}$}  \newcommand{\SffH}{S$_{\rm ffH}$} 
 \newcommand{\SH}{S$_{\rm H}$}   \newcommand{\SN}{S$_{\rm N}$} 
 \newcommand{\SL}{S$_{\rm L}$} 
 \newcommand{\SOL}{S$_{\rm oL}$} \newcommand{\SIL}{S$_{\rm iL}$} 
\newcommand{\SOF}{S$_{\rm oF}$} \newcommand{\SIF}{S$_{\rm iF}$}
\newcommand{\SOA}{S$_{\rm oA}$} \newcommand{\SIA}{S$_{\rm iA}$}
\def\ggel{ \mathrel{
   \raise1.1ex\hbox{$\scriptstyle >$}
   \mkern-10mu\raise.4ex\hbox{$\scriptstyle =$}
   \mkern-10mu\lower0.3ex\hbox{$\scriptstyle <$} }}

{\large\bf Electromagnetic extraction of energy from Kerr black holes\footnote{
Presented at ``International Conference on Theoretical Physics", 11--16 April 2005, Lebedev Physical Institute, Moscow} 
} \bigskip 

Isao Okamoto \\
{\it Institute for Black Hole Mining, 114-6 Ochikawa, Hino-shi, Tokyo 191-0034, Japan}\\
E-mail: iokamoto@m2.hinocatv.ne.jp 

\bigskip \medskip 
\hspace{2cm} \begin{minipage}[t]{13.5cm}
{\bf Abstract}\\ 
 We elucidate the `right' process for energy extraction from Kerr black holes through  `FFDE' magnetospheres, free from causality violation. It is shown that the magnetosphere of a Kerr black hole possesses the double-structure, consisting of the inner and outer magnetospheres with the pair-creation gap between them and with respective unipolar batteries at the inner and outer surfaces of the gap. 
 
 \end{minipage}

\bigskip \medskip
{\bf 1. Introduction} 

In a seminal paper Blandford \& Znajek (1977) proposed an electromagnetic mechanism of extracting rotational energy from a Kerr black hole, making use of the `boundary condition at the horizon' for the current function $I$ formulated by Znajek (1977) (referred to the BZ process or mechanism). Macdonald \& Thorne (1982) constructed the `$3+1$' formalism for general relativity in the steady axisymmetric state, which in the author's view is the most suitable one to treat black hole magnetospheres. They applied this to the BZ process, but they put the pulsar-like unipolar battery at the horizon surface. It was Phinney (1982, 1983) who first formulated MHD black hole wind theory, and Thorne et al.\ (1986) proposed the Membrane Paradigm to treat the black hole FFDE in the steady axisymmetric state. Based on the `$3+1$' formalism, Okamoto (1992) proposed the {\em effective} ergosphere and the $\al\om$ mechanism. 
\\ \hspace{1cm}
On the other hand, it were Punsly \& Coroniti (1989) that rightly pointed out that the BZ solution is mathematically proper, but physically acausal. It must indeed be pointed out that in all the papers cited above there was a seed for the later confusion lasting for two decades, that is, the `causality question'. The seed for the confusion is the `boundary conditon at the horizon' for $I$ and also the `battery at the horizon' for $\OmF$. 
\\ \hspace{1cm}
The best way to solve the `causality question' is to find the `right' process itself, which must naturally be free from claim of causality violation. To do so in FFDE, one must at first elucidate how to determine the two `force-free' constants of motion, i.e.\ $\OmF$ and $I$. This is closely related to the question where the same kind of unipolar battery as the pulsar unipolar battery should exist in the black hole magnetosphere. Just as pointed out by Punsly \& Coroniti (1989), it is obvious that any battery cannot be located at the horizon surface, which magnetic fluxes can thread, but not be frozen in. 
\\ \hspace{1cm} 
We use the force-free degenerate electrodynamics for perfectly conductive magnetospheric plasma in the steady axisymmetric state. We also adopt the `$3+1$' formalism in the Boyer-Lindquist coordinates, in which general-relativistic effects are condenced into the two metric functions $\al$ and $\om$, where $\al$ is the lapse function or redshift factor and $\om$ is the frame-dragging angular frequency. The right way leading to the `right' process is to clarify how $\al$ and $\om$ couple with the global ordered magnetic fluxes threading the horizon and extending to infinity and create the unipolar inductor(s) proper to the black hole magnetosphere. \\

{\bf 2. The key relations to the `right' process} 
 
The rotational velocity of magnetic field lines measured in the inertial frames dragged by the hole's rotation is given by 
\beeq 
\vcvF= \frac{(\OmF-\omega)\vp}{\alpha} \uvt
\label{eq:vcvF} \eneq 
where $\vp$ is the axial distance in the Boyer-Lindquist coordinates and $\uvt$ is the unit toroidal vector. Then the frozen-in poloidal electric field is 
\beeq 
\vcEp=-\frac{\vcvF}{c}\times \vcBp=-\frac{(\OmF-\omega)}{2\pi\alpha c}\vcnb \Psi 
\label{eq:vcEp} \eneq  
where $\Psi$ is the so-called stream function, and $\Psi=$constant defines a field-stream line (Macdonald \& Thorne 1982; Okamoto 1992, 2004). 
\\ \hspace{1cm} 
Eqs (\ref{eq:vcvF}) and (\ref{eq:vcEp}) possess the physical meaning of crucial significance in the black hole FFDE/MHD, because it is these relations that lead us to the `right' solution, as shown in the following (see also Okamoto 2004, 2005). The velocity of field lines $\vcvF$ plays the role of magnetic slingshot, and $\vcvF\to +\infty$ for $\vp\to\infty$ toward \Sffinf, $\vcvF=+c\uvt$ at the outer light surface \SOL, $\vcvF=0$ at the `upper null surface' \SN\ where $\om=\OmF$, $\vcvF=-c\uvt$ at the inner light surface \SIL, $\vcvF\to -\infty$ for $r\to r_{\rm H}$ toward \SffH, where \Sffinf\ and \SffH\ are the surfaces {\em near} infinity and the horizon in FFDE, bounding the force-free domains. \SN\ is a kind of {\em critical} surface where the magnetic slingshot effect changes sign, and which thereby divides the magnetosphere into the two, i.e., outer and inner, magnetospheres. In the former, $\vcvF>0$ and the magnetic slingshot effect works outwardly just as in the classical pulsar magnetosphere with $\al=1$ and $\om=0$ everywhere, whereas in the latter, $\vcvF<0$ and the magnetic slingshot effect works inwardly, that is, toward the horizon. The former survives in the classical limit of $\al=1$ and $\om=0$, with the latter vanishing. Thus the latter, i.e., the inner magnetosphere is a purely general-relativistic domain, which is produced by the $\al\om$ {\em mechanism} and referred to as the {\em effective} ergosphere (Okamoto 1992). 
Recently Levinson (2004) pointed out that spacetime around a Kerr black hole does act like a unipolar inductor through the long range gravitomagnetic force (see Beskin \& Kuznetsova 2000). Also van Putten (2004) proposes the {\em ergotube} created by the large-scale gradient of frame-dragging angular frequency coupled with magnetic fluxes along the axis of rotation. 
\\ \hspace{1cm}
The electric field in Eq (\ref{eq:vcEp}) gives rise to the potential difference between a pair of field lines (say $\Psi_1$ and $\Psi_2$), and drives the poloidal current, which leads to the Lorentz force accelerating (non-massless) particles at finite distances in MHD (with the cross-field current $\jvl>0$) or to Joule dissipation at \Sffinf\ or \SffH\ (implying a sort of substitute for MHD acceleration or increase of the hole's irreducible mass) in FFDE (with $\jpl=0$ at finite distances). \\

{\bf 3. The double-structured magnetosphere} 
 
The outer magnetosphere is separated by \SN\ from the outer one, which is crucially different from the inner one, in the sense that the magneto-{\em centrifugal} wind blows inward in the inner magnetosphere, while the wind of the same nature blows outward in the outer (see Fig.\ 
1). This inner, general-relativistic domain is produced by the $\al\om$ mechanism in the presence of global, ordered magnetic fluxes threading the horizon, and is quite different from the {\em ordinary} ergosphere defined {\em mechanically} by the static-limit surface (Okamoto 2004, 2005). 
\\ \hspace{1cm}
Just as the existence of unipolar inductor at the surface of the neutron star is responsible for driving the pulsar wind (Goldreich \& Julian 1969), so in the black hole magnetosphere some unipolar inductor(s) of the same kind must be operative to drive the black hole wind (electric current, Poynting flux, baryon-poor outflow). In addition, because the black magnetosphere must be divided to the two, the {\em dual} unipolar batteries are needed, respectively for the outgoing wind and ingoing wind. The existence of dual unipolar inductors must be related to the angular frequency of field lines $\OmF$ in some way. Then the most important issue is how to determine such a $\OmF$ as giving rise to the {\em dual} unipolar batteries in the black hole magnetosphere. This issue must be solved together with the issue of the plasma source maintaining electric current in the magnetospheres. In the pulsar's case, magnetic fluxes are frozen in the stellar matter, and $\OmF$ is usually regarded as given by the stellar angular frequency as the `boundary condition' together with charged particles of both sign (in principle), and another constant of motion $I$ is determined as the eigenvalue of the `criticality condition' at the fast surface \SF\ in terms of $\OmF$. In the black hole's case, how to determine $I$ and $\OmF$ has produced such a serious confusion as called the `causality question' as mentioned in \S 1. \\

\begin{center}
	\includegraphics[width=350pt,clip]{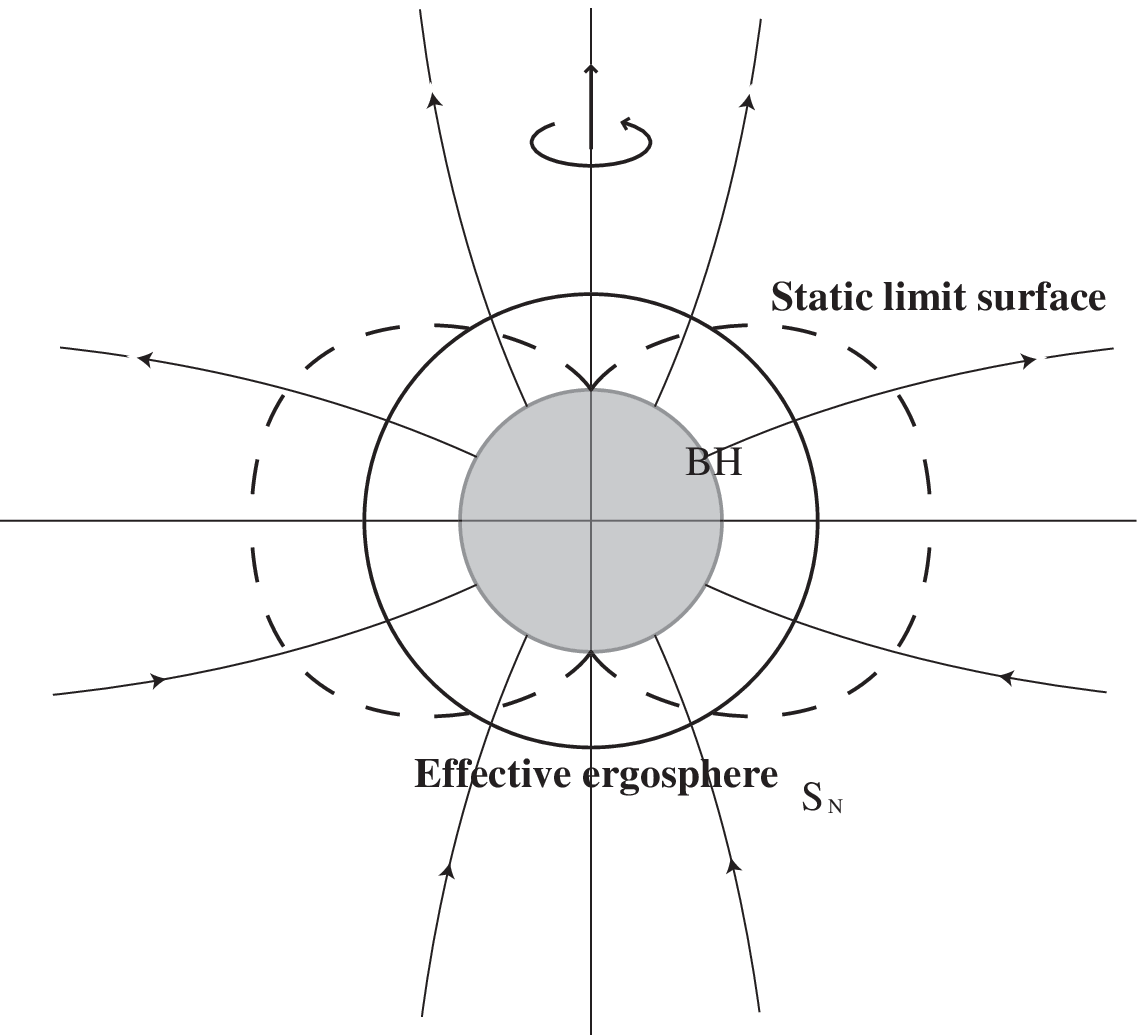}\\ \begin{minipage}[t]{13cm} {\footnotesize
{\bf Figure 1}. The double structure of a black hole magnetosphere. The outer magnetosphere is separated from the inner one by the upper null surface \SN, under which a `fine' structure with the `microphysics' is hidden in FFDE. }
\end{minipage}
\end{center}

\newcommand{\Dell}{\Delta\ell}  \newcommand{\ellN}{\ell_{\rm N}}
\newpage
{\bf 4. The dual unipolar batteries } 
 
It will be obvious that the source of charged particles and the unipolar inductors must exist between the inner and outer magnetospheres. In the force-free limit, however, it seems that \SN\ is a mere surface where there is nothing physically significant, but in reality one must consider that the pair-creation gap with the unipolar batteries at the inner and outer surfaces of it is forced to degenerate to a `thin' surface, with everything covered. To be physically consistent, one must cut the `thin' surface open in Eqs (\ref{eq:vcvF}) and (\ref{eq:vcEp}), to take out the gap with finite halfwidth $\Dell$, that is, at \SN\ where $\om(\ellN, \Psi)=\Om(\Psi)$, and one expands for $\ell=\ellN\pm\Dell$
\beeq 
\om=\OmF\mp \Delta\om, \quad  
\Delta\om=\left|\pldr{\om}{\ell}\right|_{\rm N}\Dell,
\label{eq:D/om} \eneq
where $\ell$ is measured along each field line $\Psi$ (see Fig.\ 2). From Eqs (\ref{eq:vcvF}) and (\ref{eq:vcEp}) one has 
\beeq 
\vcvF=\pm \frac{\Delta\om}{\al}\vp\uvt, \quad
\vcEp=\mp \frac{\Delta\om}{2\pi\al c}\vcnb\Psi.
\label{eq:vcvF/Ep} \eneq
It is worthwhile remarking that $\pm\Delta\om$ corresponds to $\OmF$ for the pulsar unipolar battery at the neutron star surface. Then one obtains the electromotive force (EMF) at the gap surfaces $\ell=\ellN\pm\Dell$ between a pair of field lines $\Psi_1$ and $\Psi_2$
\beeq 
{\rm EMF}_{\rm out}=-\frac{1}{2\pi c}\int^{\Psi_2}_{\Psi_1} \Delta\om d\Psi,
\label{eq:emf/out} \eneq
\beeq 
{\rm EMF}_{\rm in}=+\frac{1}{2\pi c}\int^{\Psi_2}_{\Psi_1} \Delta\om d\Psi,
\label{eq:emf/in} \eneq 
\begin{center}
	\includegraphics[width=420pt,clip]{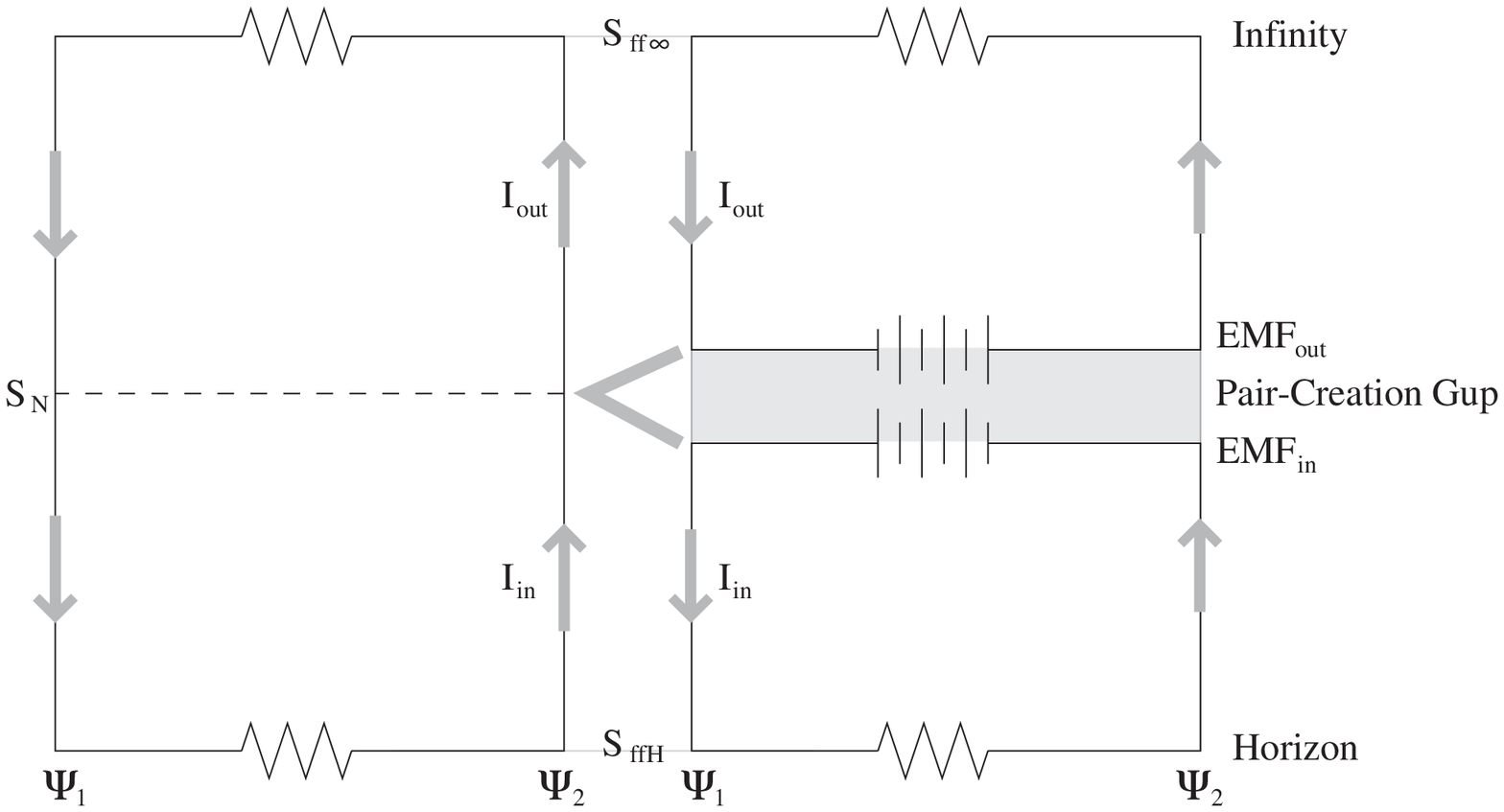}\\ \begin{minipage}[t]{13cm} {\footnotesize
{\bf Figure 2}. To reach the `right' process for extracting energy through the `force-free' magnetosphere, one must cut open the upper null surface \SN, to dig out the pair-creation gap with the dual unipolar batteries.}
\end{minipage}
\end{center}

which drive the electric currents in the outer and inner magnetospheres. The voltage drops along field lines $\Psi_1$ and $\Psi_2$ across the gap width of $2\Dell$ becomes
\beeq 
{\rm V Drop}\approx |{\rm EMF}_{\rm out}|+|{\rm EMF}_{\rm in}|\approx 2|{\rm EMF}_{\rm in}|.
\label{eq:Vdrop} \eneq 
It is this {\em stationary} voltage drop inside the gap along field lines $\Psi_1$ and $\Psi_2$ that gives rise to pair-creation and charge-separation, to input the $e^{\mp}$-flows to the outer and inner magnetopsheres and to maintain the `force-free' currents in the steady state. Then the parallel electric field $E_{\parallel}$ inside the gap is given by 
\beeq 
E_{\parallel}\approx\mp \frac{{\rm V Drop}}{2\Dell}
  \approx\mp \frac{|{\rm EMF}_{\rm in}|}{\Dell}
  \approx \mp \frac{1}{2\pi c}
         \int^{\Psi_2}_{\Psi_1} \left|\pldr{\om}{\ell}\right|_{\rm N} d\Psi
  \approx \mp \frac{(\Psi_2-\Psi_1)}{2\pi c}\left|\pldr{\om}{\ell}\right|_{\rm N} .
\label{eq:Eparallel} \eneq
where $E_{\parallel}<0$ for $\Psi_1$ and $E_{\parallel}>0$ for $\Psi_2$. This $E_{\parallel}$
will be at work continuously without being screened, to create pair-particles and charge-separate along field lines $\Psi_1$ and $\Psi_2$. Note that $E_{\parallel}$ is nearly independent of the gap width, and will survive even at the force-free limit of $\Dell\to 0$, to ensure the `current-closure condition' at \SN\ [see Eq (\ref{eq:I=I})]. \\

{\bf 5. The force-free domains } 

The `stream equation' in FFDE is given by
\beeq 
\vcnb\cdot\left[\frac{\alpha}{\vp^2}
 \left(1-\frac{\vp^2(\OmF-\omega)^2}{\alpha^2 c^2}\right)\vcnb\Psi\right]    
    + \frac{(\OmF-\omega)}{\alpha c^2}\dr{\OmF}{\Psi}|\vcnb\Psi|^2 
    + \frac{8\pi^2}{\alpha\vp^2 c^2}\dr{I^2}{\Psi}= 0
\label{eq:stream} \eneq 
in the `$3+1$' formalism (Macdonald \& Thorne 1982; Thorne et al.\ 1986; Okamoto 1992: see Blandford \& Znajel 1977 for the original expression). 
\\ \hspace{1cm}
In the above `stream equation', the `gap' is made {\em degenerated} into a `thin' surface \SN\ in the force-free limit with the gap half width $\Dell\to 0$, and every field line is so far implicitly assumed to pass through \SN\ with the constant $\OmF(\Psi)$. In reality one must presume the existence of the pair-creation gap with the dual unipolar inductors at the inner and outer surfaces of it. It is to be noted in FFDE that the poloidal electric current lines are also made {\em degenerated} to be parallel to poloidal field lines everywhere at finite distances, i.e.\ $I=I(\Psi)$, and hence no transfer of energy takes place from the field to the flow. The force-free domains are simply the one through which the electromagnetic energy  inputted flows just as the Poynting flux, and the EMF's must be given as the `boundary condition' at the boundaries and how much enegy should flow must be determined by the `criticality condition' at another boundaries, which is related to Ohm's law. This means that one cannot determine the two force-free constants of motion $\OmF$ and $I$ within the framework of force-freeness only, and hence one must solve the `double-eigenvalue problem', imposing the `criticality condition' at \SOF\ and \SIF\ and the `boundary conditon' at \SN. \\

{\bf 6. The eigenvalues for $I$} 
 
The `criticality condition' at the inner and outer fast surfaces, \SIF\ and \SOF, determines the eigenvalues for $I_{\rm in}$ and $I_{\rm out}$ as follows:
\beeq  I_{\rm in}=\frac12 (\OmH-\OmF)(B_{\rm p}\vp^2)_{\rm ffH}, \label{eq:Iin} \eneq
\beeq I_{\rm out}=\frac12 \OmF (B_{\rm p}\vp^2)_{{\rm ff}\infty} \label{eq:Iout} \eneq
(see Okamoto 2005 for details). The eigenvalues, $I_{\rm in}(\Psi)$ and $I_{\rm out}(\Psi)$, define respective ``current lines" in the inner and outer {\em eigen}-magnetospheres, along which poloidal electric currents flow. In FFDE, \SIF\ and \SOF\ are situated in the vicinity of the horizon and infinity, i.e.\ \SIF$\approx$\SffH\ and \SOF$\approx$\Sffinf, and the criticality condition there implies the termination of the force-free domains. The `current lines" cannot thread \SffH\ and \Sffinf\ and hence, deviating from corresponding field lines, must close on \SffH\ and \Sffinf\ by crossing field lines threading there. Then the `volume' currents in the force-free domains must be transformed into the `surface' currents on the `resistive' surfaces. The frozen-in electric field in Eq (\ref{eq:vcEp}) becomes 
\beeq E_{\rm p}=\frac{(\OmH-\OmF)\vp}{c}\Bp, \quad \mbox{\rm on S$_{\rm ffH}$ }, 
\eneq
\beeq E_{\rm p}=\frac{\OmF\vp}{c}\Bp, \quad \mbox{\rm on S$_{{\rm ff}\infty}$ } ,
\eneq
which combine with the eigenvalues $I_{\rm in}$ and $I_{\rm out}$, to yield Ohm's law
\beeq {\displaystyle E_{\rm p} =R_{\rm ffH}\cdot {\cal I}_{\rm ffH} }, \quad
{\displaystyle {\cal I}_{\rm ffH}
   =\left(\frac{I}{2\pi \vp}\right)_{{\rm ffH}}}, \eneq
\beeq {\displaystyle E_{\rm p} =R_{\rm ffH}\cdot {\cal I}_{\rm ffH}}, \quad 
{\displaystyle {\cal I}_{{\rm ff}\infty}
    =\left(\frac{I}{2\pi \vp}\right)_{{\rm ff}\infty}}, 
  \eneq
where
\beeq {\displaystyle R_{\rm ffH}= \displaystyle R_{{\rm ff}\infty}=\frac{4\pi}{c}=377\ 
\mbox{\rm Ohm} }. \eneq
It thus turns out that one must introduce some sort of `artificial' resistivity on \SffH\ and \Sffinf, to allow current lines to close crossing field lines threading there. Then the `Joule heating' implies the MHD acceleration beyond \SOF, and the increase of the hole's entropy beyond \SIF. \\

{\bf 7.  The eigenvalue $\OmF$} 

One has the {\em single} eigenvalue problem for the pulsar magnetosphere in the sense that $\OmF$ may be regarded as uniquely given by the angular velocity of the matter at the neutron star surface as the `boundary condition', and $I$ is determined as the eigenvalue in terms of $\OmF$ and the magnetic flux by the `criticality condition' at the fast surface \SF\ near infinity. On the other hand, the black hole has the double-structured magnetosphere consisting of the inner and outer magnetospheres, for which one has at first the respective eigenvalues of $I$ due to the `cirticality conditions' at \SIF\ and \SOF, and then these eigenvalues must be fitted by the continuity of energy and angular momentum fluxes at \SN\ as the `boundary condition', to determine the final eigenvalue of $\OmF$. That is, one has the {\em double} eigenvalue problem in the black hole magnetosphere.  
\\ \hspace{1cm} 
The expressions of energy and angular momentum fluxes flowing through the force-free black hole magnetosphere are given in terms of $\OmF$ and $I$ by
\beeq \vcS_{\rm E}=\OmF \vcS_{\rm J}
  = \frac{\OmF I}{2\pi\al c} \vcBp, \quad
 \vcS_{\rm J}=\frac{I}{2\pi\al c} \vcBp, \eneq 
(see Macdonald \& Thorne 1982, Thorne et al.\ 1986, Okamoto 1992), and then the `boundary condition' to determine the `second' eigenvalue $\OmF$ in terms of $\OmH$ is given just by 
the condition to connect the two domains, inner and outer, magnetospheres conituously, i.e.\ to the `current-closure condition' at \SN\ in FFDE  
\beeq
I_{\rm in}=I_{\rm out}. \label{eq:I=I} \eneq
Then from Eqs (\ref{eq:Iin}), (\ref{eq:Iout}) and (\ref{eq:I=I}) one has the final eigenvalue for $\OmF$: 
\beeq 
\OmF=\frac{(\Bp\vp^2)_{\rm H}} {(\Bp\vp^2)_{\rm H}+(\Bp\vp^2)_{\infty}} \OmH 
 \approx \frac12 \OmH, \label{eq:OmF/eigen}\eneq
because $(\Bp\vp^2)_{\rm H}\approx (\Bp\vp^2)_{\infty}$ in FFDE. Note the crucial difference of the condition of energy-angular momentum flux or the `current-closure condition' at \SN\ from the impedance matching of the two loads (see e.g.\ Thorne et al.\ 1986), although the final result in Eq (\ref{eq:OmF/eigen}) seems to be the same. \\

{\bf 8. Concluding remarks}  

{\bf (1)} The {\em ordinary} ergosphere is defined by the static-limit surface in the black hole `mechanics', in which the Penrose process may be at work to extract the rotational energy, but it is well known that the Penrose process itself is not so efficient for astrophysical purposes (see Phinney 1983). On the other hand, in the black hole `FFDE/MHD' the {\em effective} ergosphere is defined by the domain where $\vcvF<0$, that is, the magnetic slingshot effect works inwardly, and the ingoing magneto-{\em centrifugal} wind blows through the inner critical surfaces toward the horizon. 
\\[1.5mm] 
{\bf (2)} The {\em effective} ergosphere, i.e.\ the inner magnetosphere, is separated by the upper null surface \SN\ with $\om=\OmF$ from the classical domain, i.e.\ the outer magnetosphere. To restore the `fine' structure hidden under \SN\ in FFDE, one must expand the frame-dragging angular frequency $\om$ in the vicinity of $\om=\OmF$, with the halfwidth $\Dell$, and dig out the gap with the dual unipolar batteries at the outer and inner surfaces of it, in which charged particles are pair-created and charge-separated, to input as electric currents to the force-free domains outside. 
\\[1.5mm]
{\bf (3)} The domains outside the gap are `force-free' everywhere at finite distances, with $\vcj_{\rm p}\parallel\vcBp$ and no transfer of Poynting flux to kinetic flux takes place. One cannot determine the `force-free' constants of motion, i.e.\ $\OmF$ and $I$, within the force-free domains. These constants must be determined by breaking down the conditions of frozen-inness as well as force-freeness. 
\\[1.5mm]
{\bf (4)} The `criticality condition' at the inner and outer fast surfaces \SIF\ and \SOF\ determine the eigenvalues for $I_{\rm in}$ and $I_{\rm out}$. These combine with the frozen-in electric field, to yield Ohm's law for the surface currents on the inner and outer fast surfaces \SIF\ and \SOF, which are situated at near the horizon and infinity. This means that the force-free domains are terminated by these surfaces, and frozen-inness as well as force-freeness break down there to allow the surface currents to cross field lines threading these surfaces.
\\[1.5mm] 
{\bf (5)} The `current-closure condition' (energy conservation law) as the `boundary condition' at \SN\ determines the final eigenvalue for the angular velocity of field lines $\OmF$ and the exact location of the gap under \SN. It is however not $\OmF$ but $\Delta \om$ that appears in the expressions (\ref{eq:emf/out}) and (\ref{eq:emf/in}) for the EMF's. Thus it is the gradient of the frame-dragging $\om$ that gives rise to the dual unipolar inductors in the double-structured magnetosphere of a Kerr black hole. 
\\[1.5mm] 
{\bf (6)}  One important premise in FFDE is that magnetic field lines threading the horizon extend, passing through the pair-creation gap with dual-unipolar batteries at the two surfaces of it, to infinity, with the same angular velocity $\OmF(\Psi)$. Every `current line' must emanate from and terminate at the battery, and hence in general cannot pass through the pair-creation gap, but in the limit of $\Dell\to 0$ in FFDE with apparent disappearance of the batteries, it will be allowed to treat as if the current line crossed the gap, that is, one may impose the `current-closure condition' $I_{\rm in}=I_{\rm out}$ in FFDE. 
\\[1.5mm] 
{\bf (7)}  Kerr black holes have a double-structured magnetosphere, which consists of the classical and general-relativistic domains, and hence impose the double-eigenvalue problem for $\OmF$ and the exact location of the $\al\om$ dynamo (see Okamoto 2005 for the details). 
\\[1.5mm] 
{\bf (8)} The eigenvalue of output power in the steady {\em eigen}-state with $\OmF\approx \frac12 \OmH$ is given by
\beeq
{\cal P}=-\dr{M}{t} \approx T\dr{S}{t} \approx -\frac12 \OmH\dr{J}{t}. 
\eneq

{\bf (9)} In order to save the BZ process from causality violation, one must dig out the pair-creation gap with a pair of unipolar batteries, by cutting open the upper null surface \SN\ where $\om=\OmF\approx \frac12 \OmH$. \\

{\bf Acknowledgement} 

The author thanks the Organizing Committee of the Conference for inviting him and supporting for the local expenseses. He is also grateful to V.\ Beskin, A.\ Levinson and M.\ van Putten for discussions and comments. \\

{\bf References} 

Beskin, V.S., \& Kuznetsova, I.V., 2000, Nuovo Cimento B, 115, 795 \\
Blandford, R.D., \& Znajek, R.L., 1977, MNRAS, 179, 465 \\
Goldreich, P., \& Julian, W.H., 1969,  ApJ, 157, 869  \\
Levinson, A., 2004, ApJ, 608, 411 \\
Macdonald, D.A, \& Thorne, K., 1982, MNRAS, 198, 345 \\
Okamoto, I., 1992, MNRAS, 254, 192 \\
Okamoto, I.,   2004, Presented at the informal workshop on ``Relativistic Plasma in Magnetic \\ \hspace{3mm} Field", 16--18 August 2004, KIPAC, SLAC, Stanford \\
Okamoto, I.,   2005, MNRAS, submitted \\
Punsly, B., \& Coroniti, F., 1989, Phys.Rev.D40, 3834 \\
van Putten, M., 2004, preprint (astro-ph/0409062) \\
Znajek, R.L.,1977, MNRAS,  179, 457

\end{document}